\documentclass[a4paper,reqno,11pt]{amsart}
\pdfoutput=1

\usepackage{natbib,graphicx,multirow,amsaddr,hyperref}
\usepackage[left=1in,right=1in,top=1.4in,bottom=1.2in]{geometry}

\hypersetup{
	pdftitle={Sum of all Black-Scholes-Merton models: An efficient pricing method for spread, basket, and Asian options},
	pdfauthor={Jaehyuk Choi},
	pdfkeywords={multi-asset Black-Scholes-Merton, spread option, basket option, Asian option, curse of dimensionality},
	colorlinks=true,
	linkcolor=red,
	citecolor=blue,
	urlcolor=blue,
	pdfstartview=
}

\newcommand{\mat}[1]{\boldsymbol{#1}}
\newcommand{\matrow}[2]{\mat{#1}_{#2 *}}
\newcommand{\matcol}[2]{\mat{#1}_{* #2}}
\newcommand{\matelem}[2]{{#1}_{#2}}
\newcommand{\cov}{\Sigma}
\newcommand{\covsq}{V}
\newcommand{\wf}{g}

\newcommand{\vect}[1]{\boldsymbol{#1}}
\newcommand{\vecdot}[1]{\dot{\boldsymbol{#1}}}
\newcommand{\vecddot}[1]{\ddot{\boldsymbol{#1}}}
\newcommand{\transp}{{\scriptscriptstyle T}}
\newcommand{\distequal}{\,{\buildrel d \over =}\,}

\begin{document}

\title[Sum of all BSM models]{
	Sum of all Black-Scholes-Merton models:\\
	An efficient pricing method for\\
	spread, basket, and Asian options
}

\author[J. Choi]{Jaehyuk Choi}
\address{Peking University HSBC Business School\\
	University Town, Nanshan District, Shenzhen 518055, China}
\email{\href{mailto:jaehyuk@phbs.pku.edu.cn}{jaehyuk@phbs.pku.edu.cn}}

\thanks{The author thanks Robert Webb (editor), Jaeram Lee (discussant at the 2017 Asia-Pacific Association of Derivatives conference in Busan, Korea), and the anonymous referee for their helpful comments. This study was supported by Bridge Trust Asset Management Research Fund.}

\begin{abstract}
Contrary to the common view that exact pricing is prohibitive owing to the curse of dimensionality, this study proposes an efficient and unified method for pricing options under multivariate Black-Scholes-Merton (BSM) models, such as the basket, spread, and Asian options. The option price is expressed as a quadrature integration of analytic multi-asset BSM prices under a single Brownian motion. Then the state space is rotated in such a way that the quadrature requires much coarser nodes than it would otherwise or low varying dimensions are reduced. The accuracy and efficiency of the method is illustrated through various numerical experiments.
\end{abstract}


\maketitle

\section{Introduction}

\subsection{Background}
Ever since the celebrated success of the Black-Scholes-Merton (BSM) model, the effort to extend its simple analytic solution to the derivatives written on multiple underlying assets has been of great interest to researchers~\citep{broadie2004msoption}. One important class of such derivatives is the options on a linear combination of assets following correlated geometric Brownian motions (GBMs); these include the following three popular option types:
\begin{itemize}
	\item Spread option: European-style option on the difference between two asset prices;
	\item Basket option: European-style option on the sum of multiple asset prices with positive weights;
	\item Asian option: option on the average price of one underlying asset on a pre-determined \textit{discrete} time set or \textit{continuous} time range.
\end{itemize}
These three option types arguably represent the most actively traded non-vanilla options on exchanges or in over-the-counter markets. This is because a linear combination is a common way of associating multiple prices -- of different assets or various times -- and such options provide customized hedge or risk exposure. For examples and financial motivations, see the introductions of \citet{carmona2003spread} and \citet{linetsky2004spectral}.

However, such option pricings under the BSM model, not to mention the models beyond the BSM, is not a trivial matter. This is because, unlike normal random variables (RVs), a linear combination of correlated log-normal RVs neither falls back to the same class of distribution nor has a distribution expressed in any analytic form in general. Therefore, the exact valuation of option prices involves a multidimensional integral over positive payoffs under the risk-neutral measure. The numerical evaluation of such an integral, however, suffers from the curse of dimensionality. For example, even the coarse discretization of a standard normal distribution from $-5$ to $5$ with a grid size of $0.25$ (41 points per dimension) leads to 3 million points for four assets and 116 million points for five assets, substantially exceeding the size of a typical Monte Carlo simulation.

\subsection{Literature review\label{sec:litreview}}
A vast amount of the literature has examined each of the three option types. First, a review is conducted of the analytic methods of approximating the option price or the lower and upper bounds of the option price. For the spread option, Kirk's formula~\citep{kirk1995spread} that is widely used in practice is an approximate generalization of Margrabe's formula~\citep{margrabe1978value} for an exchange option, that is, a spread option with zero-strike price; several improvements to the formula have followed~\citep{bjerksund2014spread, lo2015pricing}. \citet{carmona2003spread} computes the lower bound of a spread option price as the maximum over the prices from all possible linear, and therefore sub-optimal, exercise boundaries. \citet{li2008spread} proposes a closed-form formula based on quadratic approximation of the exercise boundary.

Basket and Asian options share the underlying ideas for analytic approximation because the payoff of an Asian option depends on a basket of correlated prices over the observation period. One of the most popular approaches is to approximate the distribution of the GBM sum with other analytically known distributions, such as log-normal~\citep{levy1992average,levy1992avgccy}, reciprocal gamma~\citep{milevsky1998asian,milevsky1998basket}, shifted log-normal~\citep{borovkova2007closed}, and log-extended-skew-normal~\citep{zhouwang2008} distributions, and with perturbation expansions from known distributions~\citep{turnbull1991quick,ju2002}.
Another popular idea is to exploit the geometric mean of GBMs---as opposed to arithmetic mean---whose distribution is log-normal and hence analytically solvable. The option price on the geometric mean is a reasonable proxy for that on the arithmetic mean~\citep{gentle1993basket}. Thus, it can be used as a control variate, thereby reducing the Monte Carlo variance for Asian~\citep{kemna1990pricing} and basket~\citep{wilmott_basket} options. \citet{curran1994valuing} uses the geometric mean as a conditioning variable to analytically estimate option prices. The conditioning approach is further refined by \citet{beisser1999another} and \citet{deelstra2004basket} and also applied to the continuously monitored Asian options \citep{rogers1995value}. For other basket and Asian option pricing approaches as well as classifications, see \citet{zhouwang2008} and the references therein.

Analytic approximation methods are appealing because of their simple computation, but has a major limitation in that the results are not precise. While each method is accurate for certain parameter ranges where the underlying assumptions are valid, the accuracy of any one method can hardly be validated for all ranges. For example, no single method performs well in basket options under various parameter sets \citep{wilmott_basket}, although the method of \citet{ju2002} is outstanding overall.
Therefore, practitioners must carefully identify the parameter range in which the method of interest works best. Given the multi-asset aspect of such problems, the \textit{charting} of parameter maps in advance is not a trivial matter. Moreover, since errors cannot be controlled in approximation methods, the tendency is to eventually apply external methods, typically a Monte Carlo simulation, to obtain a benchmark value.

Convergent pricing methods are fewer in number compared to approximation methods. \textit{Convergent} implies that the method can produce a deterministic price, as opposed to Monte-Carlo methods, and converge to the true value with a reasonable amount of computation when the computational parameters (e.g., grid size) are tuned. Convergent methods are feasible for spread options because the problem is two-dimensional. \citet{ravindran1993lowfat} and \citet{pearson1995spread} reduce the pricing problem to a one-dimensional integration over the BSM prices with varying spot and strike prices. In addition, \citet{dempster2002spread} and \citet{hurdzhou2010spread} apply a two-dimensional fast Fourier transform (FFT). 

To the best of the author's knowledge, few studies on basket options use methods that can be defined as convergent. Particularly, no attempt has been made to use direct integration, even for error measurement, thus indicating the challenge posed by the approach (see \S~\ref{sec:example}). Although the FFT approach proposed by \citet{leentvaar2008multi} reduces the computation time through parallel partitioning, it does not significantly reduce the computation amount.

Previous convergent methods used for Asian options are based on the idea that pricing involves a single price process over time. The continuously averaged Asian option, although hardly traded in practice owing to contractual difficulty, has analytic solutions comprising the triple integral~\citep{yor1992some}, Laplace transform in maturity~\citep{geman1993bessel}, and a series expansion~\citep{linetsky2004spectral}. For discrete averaging, a series of studies have exploited the recursive convolution, referred to as the Carverhill-Clewlow-Hodges factorization, on the probability density function (PDF) or price \citep{carverhill1990risk,benhamou2000fast,fusai2008pricing,cerny2011asian,fusai2011pricing,zhang2013efficient}. \citet{cai2013asian} describes the price of a discretely monitored Asian option as an asymptotic expansion on a small observation interval.

Despite structural similarity, only a few studies are applicable to all the three option types. \citet{carmona2005multivariate} extend the lower bound approach~\citep{carmona2003spread} to a multi-asset problem, while \citet{deelstra2010pricing} use the commonality theory to approximate the option price. However, because these studies follow the analytic approximation class, their methods have the aforementioned limitations. Essentially, no convergent method can work consistently for all the three option types.

\subsection{Contribution of paper}
The aim of this study is to find an efficient and unified pricing method for options on the linear combinations of correlated GBMs. While the common assumption is that outright multidimensional integration is computationally prohibitive, an innovative integration scheme is developed, which can significantly alleviate the curse of dimensionality.
First, it is observed that if all the price processes are driven by a single Brownian motion, the option price can be analytically obtained by a multidimensional extension of the BSM formula, with the exercise boundary obtained from a numerical root-finding. Therefore, the option price can be integrated analytically for the first dimension and numerically for the remaining dimensions. The key to the current approach is to choose the first dimension via factor rotation such that the analytic price from the first dimension behaves well (i.e., smooth and slowly varying) as an integrand for the numerical integration that follows. Numerical experiments show that even a coarse discretization produces a very accurate price, and that the price quickly converges to the true value as the number of nodes increase.

The contributions of this study in the context of each option type are presented in order below.
Here, the approach to spread options is similar to those of \citet{ravindran1993lowfat} and \citet{pearson1995spread}; that is, the integration in the first dimension is done analytically. However, the factor rotation in this study's method significantly reduces the cost of numerical integration in the second dimension. Even at the expense of extra computation due to numerical root-finding, which is not found in the aforementioned studies, the overall computation is much lighter owing to reduced discretization on the second dimension. 

As regards basket options, the method used here is fully convergent. While the speed of convergence depends on the covariance structure, in general, the method can converge to the true option value for a wide range of parameters and dimensions. For the first time, the converged prices for several benchmark tests used in the literature can be reported.

As for the study on Asian options, this study's pricing scheme serves as a novel alternative to those of previous works. The integration approach is particularly problematic for Asian options owing to large dimensionality, that is, a large number of observations.
In effect, the study utilizes several of the first factors of a Brownian motion series representation as with the principal component analysis (PCA). Therefore, the method falls short of being truly convergent. However, the method is accurate for all practical purposes and enables cheaper computation compared to existing methods. Furthermore, it is capable of pricing a continuously monitored Asian option in a discrete monitoring framework, which is a rare approach compared to those in the opposite direction. To this extent, the method can be used to flexibly handle features such as non-uniform weights, non-uniform averaging intervals (e.g., forward-start averaging), and time-dependent volatility, which are difficult to incorporate in methods based on the continuum theory~\citep{linetsky2004spectral,cai2013asian,fusai2011pricing}. Asian options are discussed in detail in \S~\ref{sec:asianopt}.

This study focuses on the BSM model, but its findings can be useful for other models too. The results can be trivially modified to apply to displaced GBMs, as shown in a numerical example in \S~\ref{sec:result}. Displaced GBMs have an extra degree of freedom to capture volatility skew, if not full smile, as observed in the option market. 
The results can also be applied to stochastic volatility models, such as the \citet{hullwhite1987sv}, \citet{heston1993closed}, and stochastic-alpha-beta-rho \citep{hagan2002managing} models. In these models, the options are priced using the conditional Monte-Carlo method, where the price is expressed as an expectation of the BSM prices, conditional on quantities such as terminal volatility and integrated variance; see \citet{willard1997condmc}, \citet{broadie2006mcheston}, and \citet{cai2017sabr}, respectively. Thus, an efficient BSM pricing method is critical for pricing the spread or basket options under such approaches. However, a detailed implementation is beyond the scope of this study.

This paper is organized as follows. Section~\ref{sec:setup} formulates the problem. Section~\ref{sec:integration} outlines the multidimensional integration scheme. Section~\ref{sec:pfactor} discusses the optimal rotation of the dimensions facilitating integration. Section~\ref{sec:asianopt} discusses the implications for Asian options. Section~\ref{sec:result} reports the numerical results. Finally, Section~\ref{sec:conc} concludes the study.

\section{Model setup and preliminaries\label{sec:setup}}
Assume that asset prices, $S_k$ for $1\le k \le N$, follow the correlated GBMs under risk-neutral measure
\begin{equation}
\label{eq:sde}
\frac{d S_k(t)}{S_k(t)} = (r-q_k)\,dt + \sigma_k \, dW_k(t),
\end{equation}
where $\sigma_k$ is the volatility, $q_k$ is the dividend rate, $r$ is the risk-free interest rate, and $W_k(t)$ is a standard Brownian motion with correlation $dW_k(t) dW_j(t) = \rho_{kj} dt$ ($\rho_{kk}=1$).
The final payoff of the options considered here depends on a linear combination of the asset prices observed earlier than or at the expiry $T$. The payoff of a vanilla call option with strike price $K$ is as follows:
\begin{equation}
\Big( \sum_{k=1}^N w_k\,S_k(t_k) - K \Big)^+,
\end{equation}
for weights, $w_k$, and observation times, $0\le t_k\le T$. Here, $(x)^+=\max(x,0)$ is the positive-part operator. The current setup is generic enough to include, but not limited to, the following three option types:
\begin{itemize}
	\item European spread option: $w_k<0$ for some, but not all, $k$; additionally, $t_k=T$ for all $k$.
	\item European basket option: $w_k>0$ and $t_k=T$ for all $k$.
	\item Asian option: $w_k>0$ for all $k$ with $\sum w_k=1$ and $0 \le t_1<\cdots<t_N = T$. The price processes are all identical, $S_k(t) = S_j(t)$ for $k\neq j$; thus, the index $k$ is omitted without ambiguity; for example, $S(t)$, $W(t)$, $\sigma$, and $q$. The continuously monitored Asian option,
	whose payoff is $(\frac1T \int_0^T S(t) dt - K)^+$, will be considered under the discrete framework.
\end{itemize}

Next, a few notations and conventions are set. For matrix $\mat{A}$, the following is noted: the $k$-th row vector of $\mat{A}$, the $j$-th column vector of $\mat{A}$, and the $(k,j)$ component of $\mat{A}$ by $\matrow{A}{k}$, $\matcol{A}{j}$, and $\matelem{A}{kj}$, respectively. Using these notations, for example, the matrix multiplication $\mat{C} = \mat{A}\mat{B}$ can be expressed as $\matelem{C}{kj} = \matrow{A}{k}\matcol{B}{j}$. The transpose of $\mat{A}$ is noted by $\mat{A}^\transp$ and the identity matrix by $\mat{I}$. For vector $\vect{x}$, the $k$-th component is noted by $x_k$. Moreover, unless otherwise stated, the vector is a column vector. The $L^2$-norm of $\vect{x}$ is $|\vect{x}| = \sqrt{\vect{x}^\transp\vect{x}}$ and the Frobenius norm of $\mat{A}$ is $\| \mat{A} \|_F = (\sum_{k,j} \matelem{A}{kj}^2)^{1/2}$. Further, unless otherwise specified, the dimensions of a matrix are $N\times N$, the size of a vector is $N$, and the indices, $k$ and $j$, run from $1$ to $N$. For the factor matrix $\mat{\covsq}$ to be defined below, $k$ is used for indexing rows (assets) and $j$ for indexing columns (Brownian motions or factors). Throughout this study, the terms \textit{factor} and \textit{dimension} are used interchangeably.

Under the BSM model, the log prices, $\log S_k(t_k)$, follow correlated normal distributions, with
the covariance matrix $\mat{\cov}$ given as
\begin{equation}
\label{eq:def_cov}
\matelem{\cov}{kj} = \rho_{kj} \sigma_k \sigma_j \min(t_k, t_j).
\end{equation}
If $\mat{\covsq}$ is a square root matrix of $\mat{\cov}$, satisfying $\mat{\covsq} \mat{\covsq}^\transp = \mat{\cov}$, then the observation $S_k(t_k)$ can be decorrelated to
\begin{equation}
S_k(t_k) \distequal F_{k} \exp(-\frac12 \matelem{\cov}{kk}  + \matrow{\covsq}{k} \vect{z} ),
\end{equation}
where $\vect{z}$ is a vector of independent standard normal RVs and $F_k = S_k(0)\, e^{(r-q_k) t_k}$ is the $t_k$-forward price observed at $t=0$. The symbol $\distequal$ stands for equality in distribution law.
Since $\mat{V}$ is multiplied to the state vector $\vect{z}$, it is referred to as a risk factor matrix, or simply a factor matrix. The square root matrix $\mat{V}$ is not unique.
Although the Cholesky decomposition $\mat{C}$ is a popular choice, any matrix rotated from $\mat{C}$, such as $\mat{\covsq} = \mat{C} \mat{Q}$ for an orthonormal matrix $\mat{Q}$, is also a square root matrix of $\mat{\cov}$. However, note that the norm of row vectors $|\matrow{V}{k}|$ is invariant under any rotation; since $|\matrow{V}{k}|^2$ is the variance of the $k$-th asset's return,
$|\matrow{V}{k}|^2 = \matrow{V}{k}\matcol{V}{k} = \matelem{\cov}{kk}$.
Further, the Frobenius norm of $\mat{\covsq}$ is also invariant because $\| \mat{V} \|_F^2 = \sum_k |\matrow{\covsq}{k}|^2 = \sum_k \matelem{\cov}{kk}$.

The forward value of the call option price becomes an $N$-dimensional integration,
\begin{equation} \label{eq:int_Cn}
C = \int_{\vect{z}} \Big( \sum_k w_k F_k \exp(-\frac12 \matelem{\cov}{kk}  + \matrow{\covsq}{k} \vect{z} ) - K \Big)^+ n(\vect{z})\, d\vect{z},
\end{equation}
where $n(\vect{z})$ is a multivariate standard normal PDF.

\section{Integration scheme\label{sec:integration}}
\subsection{Single-factor BSM formula on first dimension}
The BSM model's analytic tractability can be extended to the multi-asset case when a single Brownian motion drives all assets. Consider the integration over the first dimension $z_1$ only:
\begin{equation}
\label{eq:int_C1}
C_\text{BS}(\vecdot{z}) = \int_{-\infty}^\infty	\Big( \sum_k w_k F_k\, f_k(\vecdot{z}) \exp\big(-\frac12\matelem{\covsq}{k1}^2 + \matelem{\covsq}{k1} z_1\big) - K \Big)^+ n(z_1)\, dz_1,
\end{equation}
where the dependence on other dimensions is absorbed into the coefficient function defined as
\begin{equation}
f_k (\vecdot{z}) = \exp\Big(-\frac12 \sum_{j=2}^{N}\matelem{\covsq}{kj}^2 + \matrow{\covsq}{k} \vecdot{z} \; \Big)
\quad\text{for}\quad \vecdot{z} = (0,z_2, \cdots, z_N)^\transp.
\end{equation}
While $\vecdot{z}$ has $z_1=0$ for computational convenience, $F(\vecdot{z})$ for function $F$ should be understood as $F(z_2, \cdots, z_N)$.
Note that $f_k(\vecdot{z}) > 0$ and $E(f_k(\vecdot{z})) = 1$ for all $k$. The value $C_\text{BS}(\vecdot{z})$ can be seen as the price of an option on the weighted asset prices driven by a single Brownian motion, where the forward price is $F_k\, f_k(\vecdot{z})$ and the standard deviation of the log price is $\matelem{\covsq}{k1}$ for the $k$-th asset.

This one-dimensional integration can be analytically evaluated only if the region of the positive payoff is identified. However, to find the roots of the payoff, a numerical method must be used, like the Newton-Raphson method. 
As explained below in \S~\ref{sec:factor1}, $\mat{\covsq}$ can be chosen such that the payoff monotonically increases in $z_1$ and there always exists a unique root, $z_1=-d(\vecdot{z})$, of the equation
\begin{equation}
\label{eq:rootn}
\sum_k w_k F_k\, f_k(\vecdot{z}) \exp\big(-\frac12\matelem{\covsq}{k1}^2 + \matelem{\covsq}{k1} z_1\big) = K.
\end{equation}
The integration from $-d(\vecdot{z})$ to $\infty$ yields
\begin{equation} \label{eq:C_BS}
C_\text{BS}(\vecdot{z}) = \sum_{k} w_k F_k f_k(\vecdot{z}) N(d(\vecdot{z})+\matelem{V}{k1}) - K N(d(\vecdot{z})).
\end{equation}
This is a multi-asset extension of the BSM formula. The original BSM formula is a special case of the single asset case ($N=1$, $t_1=T$, and $w_1=1$):
\begin{equation}
C = F_1\, N(d+\matelem{V}{11}) - K\, N(d) \quad \text{for} \quad
d = \frac{\log(F_1/K)}{\matelem{\covsq}{11}} - \frac{\matelem{\covsq}{11}}{2},
\end{equation}
where $\mat{\covsq}$ is a scalar value, $\matelem{V}{11} = \sqrt{\matelem{\cov}{11}} = \sigma_1 \sqrt{T}$.

Despite the cumbersome numerical root-finding, the analytic integration of $z_1$ plays an important role, which involves more than simply reducing one dimension in the integration. First, because of the cusp of the option payoff at the strike, numerical integration on the first factor would have required the densest discretization. Thus, analytic integration allows us to skip the most computationally costly dimension, albeit at the expense of numerical root-finding. Second, because the first dimension has a certain degree of freedom from factor matrix rotation, $\mat{\covsq}$ can be chosen in favor of the numerical integrations that follow (see \S~\ref{sec:factor1}). The integration on the first factor is precise and computationally inexpensive, regardless of the choice of $\mat{\covsq}$, because it is analytic.
Last, analytic pricing can capture the tail probability (e.g., option price of a far-out-of-the-money strike), which Monte Carlo simulation or discretization-based numerical integration cannot easily do.

\subsection{Quadrature integration on other dimensions}
Integration over other dimensions $\vecdot{z}$ can be performed using a numerical quadrature. As the integration is weighted by normal distribution density, Gauss-Hermite quadrature (GHQ) is the most suitable choice. Let $\{\vecdot{z}_m\}$ and $\{h_m\}$ be the points and weights, respectively, of the GHQ associated with $n(\vecdot{z})$, generated over the dimensions, $(z_2,\cdots,z_N)$. Subsequently, the option price becomes a weighted sum as follows:
\begin{equation}
\label{eq:call_quad}
C = \int_{\vecdot{z}} C_\text{BS}(\vecdot{z})\,n(\vecdot{z})\, d\vecdot{z} = \sum_{m=1}^{M} \;h_m\;  C_\text{BS}(\vecdot{z}_m).
\end{equation}
Here, $M$ is the total number of nodes, $M=\prod_{j=2}^{N} M_j$, where $M_j$ is the node size of the $j$-th dimension. Therefore, the option price is casted on a linear combination of asset prices into the weighted sum of the multi-asset BSM prices in (\ref{eq:C_BS}), where the forward prices of the assets vary as $F_k\, f_k(\vecdot{z}_m)$.

Moreover, the same integration scheme can be applied to the quantities of interest other than the call option price.
A few examples are given below.
\begin{itemize}
	\item \textbf{Price of put option}:
	\begin{equation} \label{eq:put_quad}
	P = \sum_{m=1}^{M} \;h_m\; P_\text{BS}(\vecdot{z}_m)\quad\text{where}\quad P_\text{BS}(\vecdot{z})= K N(-d(\vecdot{z})) - \sum_k w_k F_k f_k(\vecdot{z}) N(-d(\vecdot{z})-\matelem{V}{k1}).
	\end{equation}
	
	\item \textbf{Price of binary call option}:
	\begin{equation}
	D = \sum_{m=1}^{M} h_m N(d(\vecdot{z}_m)).
	\end{equation}
	
	\item \textbf{Delta of forward price $F_k$}:
	\begin{equation} \label{eq:delta}
	D_k = w_k \sum_{m=1}^{M} h_m f_k(\vecdot{z}_m)\,N(d(\vecdot{z}_m) + \matelem{\covsq}{k1}).
	\end{equation}
\end{itemize}

\subsection{Dimensionality reduction\label{sec:reduction}}
Since the problem involves the factor matrix $\mat{\covsq}$, it naturally leads to the possible dimensionality reduction of low varying factors in the context of PCA. To minimize the side effects, the following approach is adopted. Without loss of generality, assume that the factor strengths $|\matcol{\covsq}{j}|$ are in decreasing order of $j$ and the aim is to reduce the dimensions of $N'<j\le N$ because these $|\matcol{\covsq}{j}|$ are small. Thus, it is assumed that the variation of $d(\vecdot{z})$ on such $z_j$ is also small and the dependence on these dimensions is ignored as $d(\vecdot{z}) \approx d(\vecddot{z})$, where $\vecddot{z} = (0,\,z_2, \cdots, z_{N'},\, 0, \cdots)^\transp$ is the state vector of the surviving dimensions, with zeros padded to the rest for convenience. In addition, notation $d(\vecddot{z})$ should be similarly interpreted as $d(z_2,\cdots,z_{N'})$.
Because the dependence of $C_\text{BS}(\vecdot{z})$ on the reduced dimensions occurs only through $f_k(\vecdot{z})$, the integration of (\ref{eq:call_quad}) over the truncated dimensions can be moved to (\ref{eq:C_BS}) instead, which can be done analytically. With abuse of notation, the integral of $f_k (\vecdot{z})$ over those dimensions can similarly be given as
\begin{equation}
f_k (\vecddot{z}) = \exp\big(-\frac12 \sum_{j=2}^{N'}\matelem{\covsq}{kj}^2  + \matrow{\covsq}{k} \vecddot{z} \;\big).
\end{equation}
Thus, the forward price can be preserved as $F_k = E(F_k\,f_k(\vecddot{z}))$, even after dimensionality reduction.
The previous results---(\ref{eq:rootn}), (\ref{eq:C_BS}), and (\ref{eq:call_quad})---remain remarkably consistent under reduced dimensions through pure notational changes---$N$ to $N'$, $\vecdot{z}$ to $\vecddot{z}$, and $f_k(\vecdot{z})$ to $f_k(\vecddot{z})$.
In particular, through quadrature integration (\ref{eq:call_quad}) over $\vecddot{z}$, the total number of nodes is reduced to $M=\prod_{j=2}^{N'} M_j$. This dimensionality reduction is critical for pricing Asian options, as discussed later in \S~\ref{sec:asianopt}.

\subsection{Forward price as control variate\label{sec:CV}}
In case of too sparse quadrature nodes, the error from integration can be non-negligible. This error can be reduced by using the forward price $F_k$ as control variate. Let $\bar{f_k}$ be the numerically evaluated expectation of $f_k(\vecdot{z})$, $\bar{f_k} = \sum_{m=1}^M h_m \,f_k(\vecdot{z}_m)$, which is not exactly equal to 1. For example, for a standard normal $z$, $E(e^{-\frac12+z})$ deviates by $-6.3\times 10^{-3}$ under the GHQ evaluation with three nodes; it also deviates by $-4.6\times10^{-4}$ with four nodes from the true value $1$.
Thus, $F_k$ is mispriced by $F_k(\bar{f_k}-1)$ under the GHQ evaluation. This error is also present in the put-call parity of (\ref{eq:call_quad}) and (\ref{eq:put_quad}):
\begin{equation}
C - P - \sum_k (w_k\,F_k) + K = \sum_k w_k\,F_k(\bar{f_k} - 1).
\end{equation}
This leads to problems such as inconsistent implied volatilities for put and call options with the same strike price. Since the sensitivity (delta) of each $F_k$ is computed as (\ref{eq:delta}), the option prices can be adjusted by using them as the coefficients of the control variate:
\begin{equation}
C' = C - \sum_k D_k F_k(\bar{f_k} - 1), \quad
P' = P - \sum_k (D_k - w_k) F_k(\bar{f_k} - 1).
\end{equation}
The put-call parity of the adjusted prices, $C'$ and $P'$, holds exactly.

\section{Optimal choice of risk factor matrix\label{sec:pfactor}}
\subsection{Illustrative example} \label{sec:example}
\begin{figure}
	\centering
	\caption{Pricing of the basket put option with payoff
	$(K-e^{x_1} - e^{x_2})^+$ for uncorrelated standard normals $x_1$ and $x_2$.
	(a) shows the exercise region for $K=2$ and $4$ (shaded area) along with the boundary $-d(x_2)$ (solid line). The axis $(z_2,z_1)$ is a $45^\circ$ degree rotation from the original axis $(x_2,x_1)$. (b) shows the integration along the first dimension of two coordinate systems, $x_1$ (solid blue) and $z_1$ (dashed red).
	(c) shows the error of the GHQ integration of the functions in (b) under $(x_2,x_1)$ (diamond) and  $(z_2,z_1)$ (circle) for increasing node size.
	}  \label{fig:bdd}
	\begin{minipage}{0.45\linewidth}
		(a)\\ 
		\includegraphics[width=\textwidth]{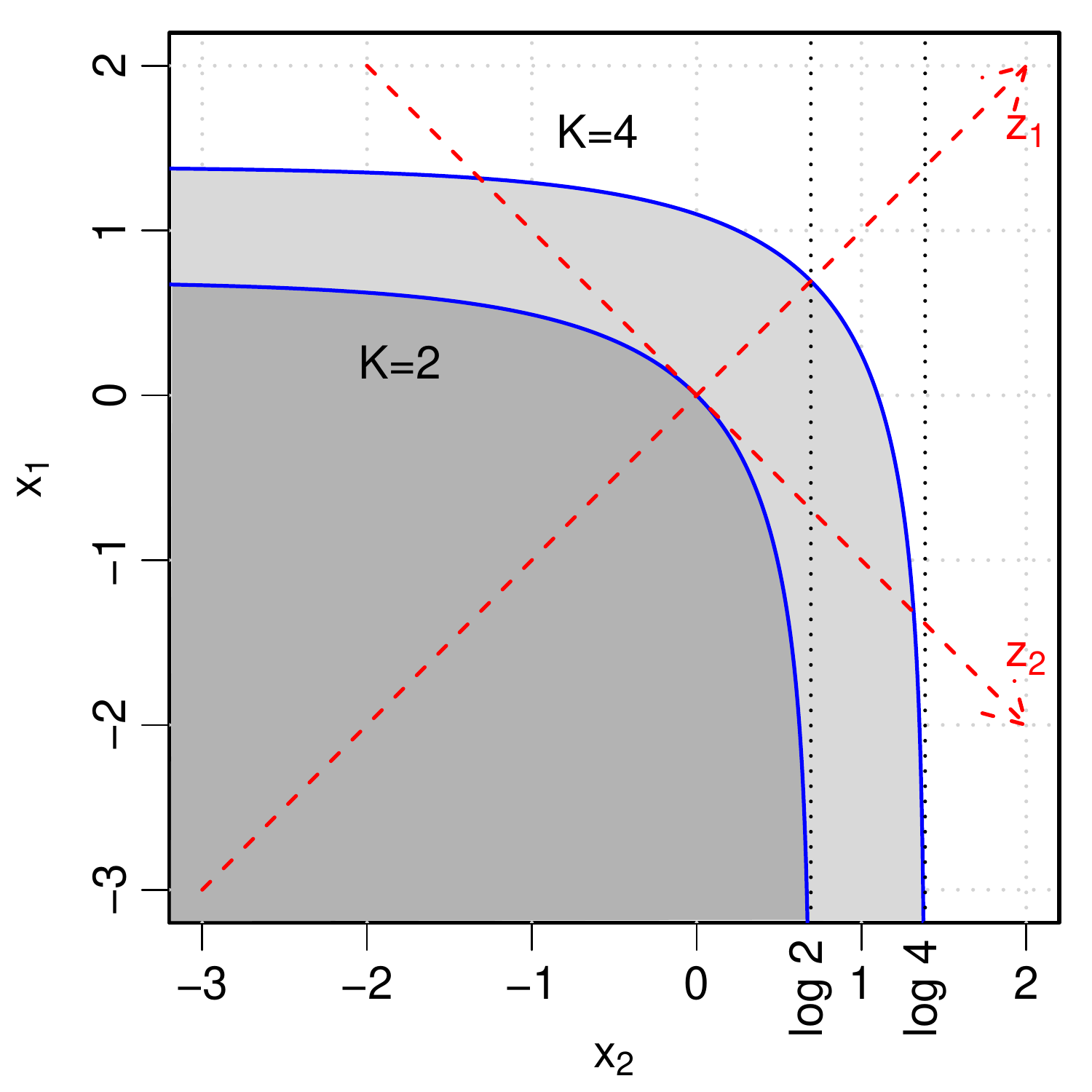}
	\end{minipage}\\
	\begin{minipage}{0.45\linewidth}
		(b)\\
		\includegraphics[width=\textwidth]{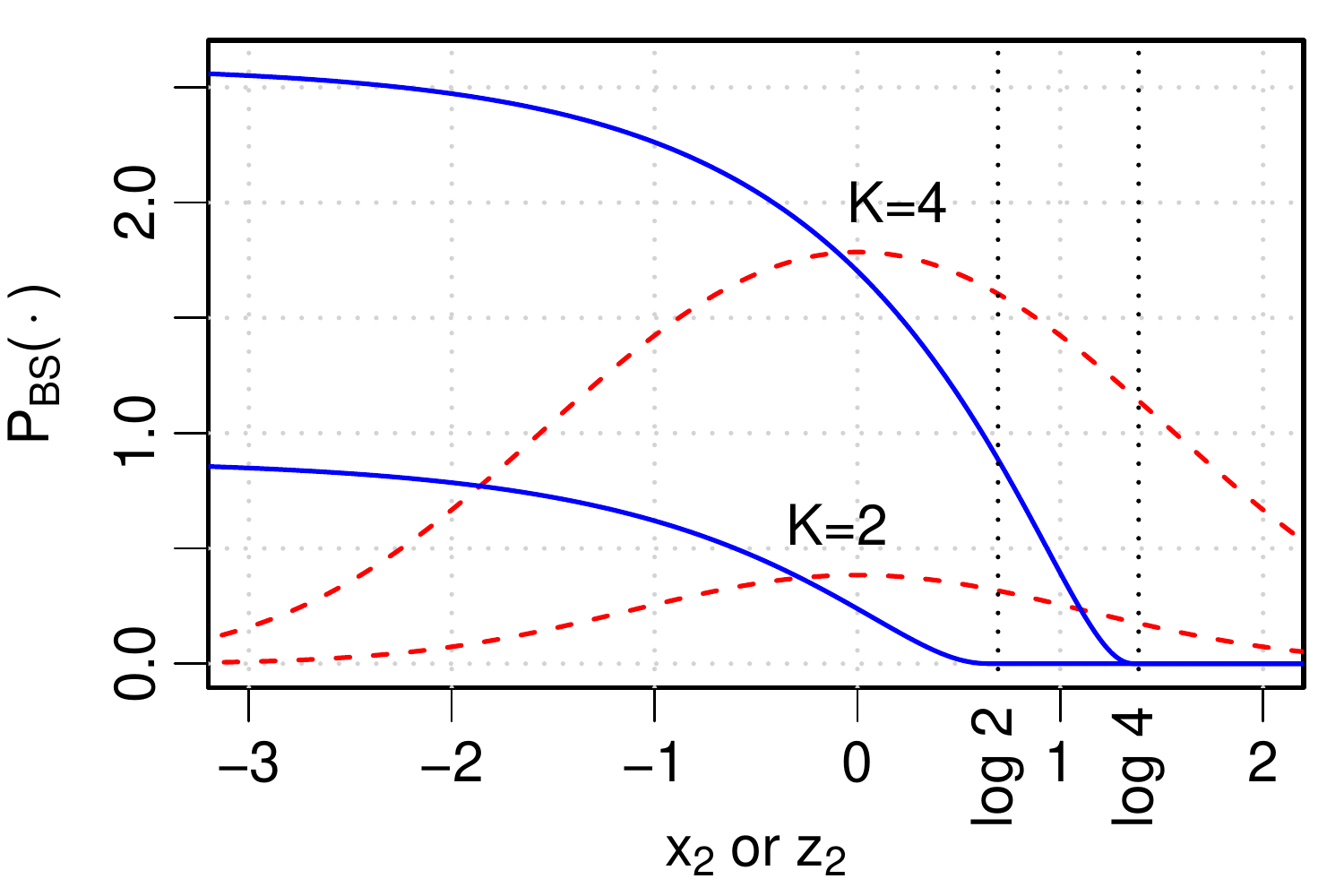}
	\end{minipage}\\
	\begin{minipage}{0.45\linewidth}
		(c)\\
		\includegraphics[width=\textwidth]{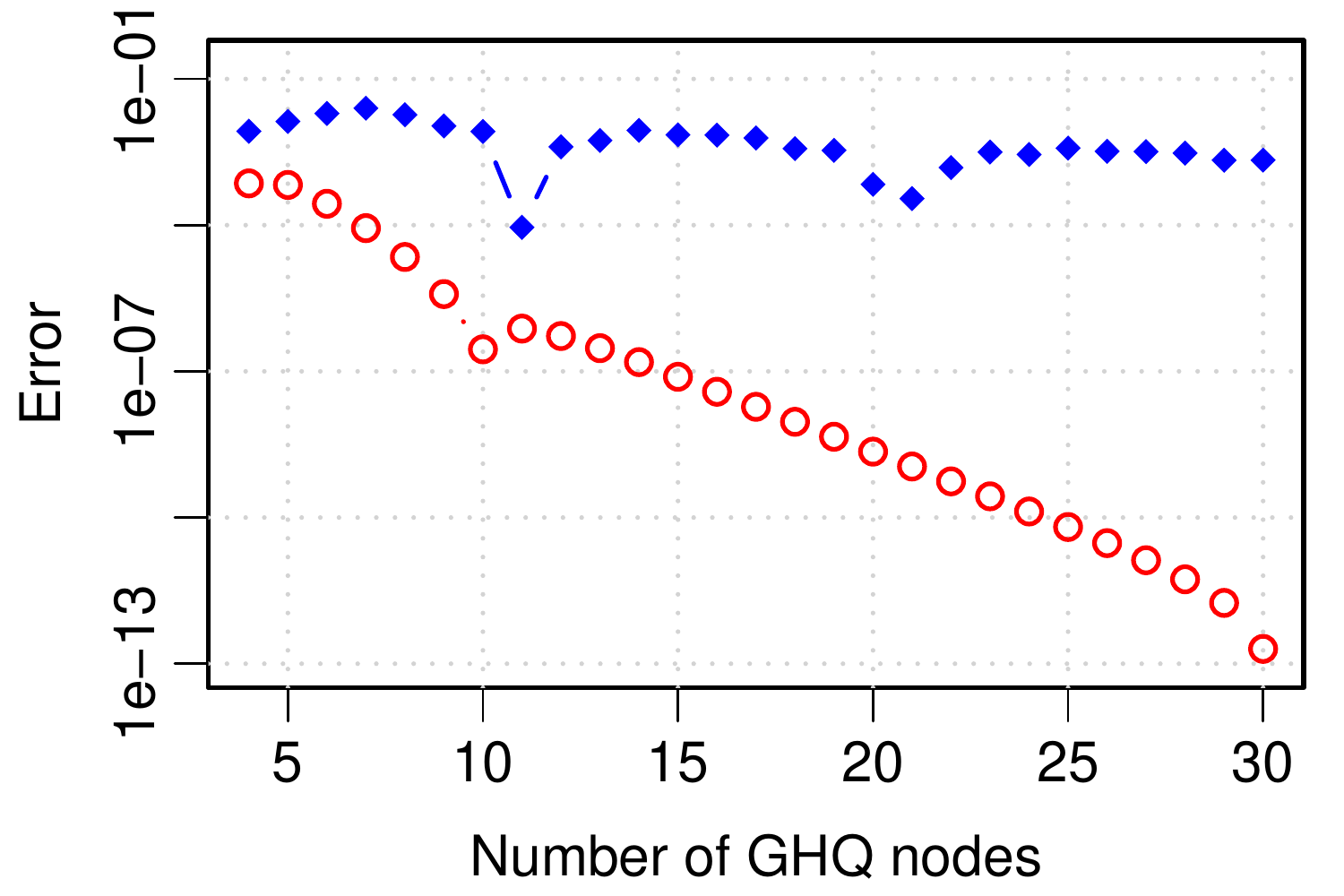}
	\end{minipage}
\end{figure}

Through a simple example, it is demonstrated why a naive numerical integration suffers from slow convergence and how a proper rotation of the factor matrix can improve convergence. Consider a basket put option on two uncorrelated assets ($N=2$) with parameters $\mat{\cov}=\mat{I}, r=0, w_k=1, F_k=e^{1/2}$ and $q_k=0$ for $k=1, 2$ to ensure that the payoff is given as $(K-e^{x_1} - e^{x_2})^+$ for the independent standard normal RVs, $x_1$ and $x_2$. A put option is chosen to clearly illustrate the singularity from the vanishing exercise region. However, the same result holds for the call option also owing to put-call parity. The put option price is
\begin{equation}
P = \int P_\text{BS}(x_2) dx_2 \quad\text{for}\quad
P_\text{BS}(x_2) = (K-e^{x_2}) N(-d(x_2)) - e^{\frac12} N(-d(x_2)-1),
\end{equation}
where $P_\text{BS}(x_2)$ is the integration of the payoff along the $x_1$ axis from $x_1=-\infty$ to $-d(x_2) = \log(K - e^{x_2})$. As shown in Fig~\ref{fig:bdd}(a), the exercise boundary $-d(x_2)$ diverges to $-\infty$ as $x_2$ approaches $\log(K)$ from the left; thus, $P_\text{BS}(x_2) = 0$ for $x_2\ge\log(K)$.
In order to accurately evaluate the numerical integration over the $x_2$ axis, the discretization around the singularity at $x_2 = \log(K)$ should be dense.

Alternatively, consider the $45^\circ$-rotated coordinate $(z_1,z_2)$ under which the payoff becomes $( K-e^{(z_1-z_2)/\sqrt{2}} - e^{(z_1+z_2)/\sqrt{2}})^+$. This yields the exercise boundary, $z_1 = -d(z_2) = -\sqrt{2}\log\big(2\cosh(z_2/\sqrt2)/K \big)$, and the option price becomes
\begin{equation}
P = \int P_\text{BS}(z_2) dz_2 \quad\text{for}\quad
P_\text{BS}(z_2) = K\,N(-d(z_2)) - 2e^{\frac14}\cosh(\frac{z_2}{\sqrt{2}}) N(-d(z_2)-\frac{1}{\sqrt{2}}).
\end{equation}
Since the boundary $-d(z_2)$ exists at all $z_2$, $P_\text{BS}(z_2)$ is infinitely differentiable in all $z_2$ and hence suitable for numerical integration along $z_2$ (see Fig~\ref{fig:bdd}(b) for $P_\text{BS}(x_2)$ and $P_\text{BS}(z_2)$.) As shown in Fig~\ref{fig:bdd}(c), the error from quadrature integration under $(z_2,z_1)$ decreases exponentially as the number of nodes increases, but the error under $(x_2,x_1)$ decreases very slowly.

\subsection{Selection of first factor\label{sec:factor1}}
The aforementioned intuition is generalized to the correlated and higher dimensional cases. 
The following two criteria are set for the selection of $\mat{V}$: (i) the exercise boundary $d(\vecdot{z})$ of (\ref{eq:rootn}) should exist for all $\vecdot{z}$ and $K$, and (ii) the variation of $d(\vecdot{z})$, $|\partial d(\vecdot{z})/\partial z_j|$, for $j\ge 2$, should be minimized.
The purpose is to make $C_\text{BS}(\vecdot{z})$ not only differentiable over $\vecdot{z}$, without $-d(\vecdot{z})$ diverging, but also low varying to the maximum extent possible.

As the coefficient functions $\{f_k(\vecdot{z})\}$ can take almost any arbitrary positive values, $w_k \matelem{V}{k1}>0$ is imposed for all $k$ as a sufficient condition to satisfy (i).
In such a constraint, the left-hand side of (\ref{eq:rootn}) represents a strictly monotonic function of $z_1$, with the value range of $(0,\infty)$ for basket and Asian options or $(-\infty,\infty)$ for spread options. Hence, a unique root $d(\vecdot{z})$ exists for any $\vecdot{z}$ and non-trivial $K$.

For (ii), the following linearized approximation of the GBM is applied: $\exp(-\frac12 \matelem{V}{kj}^2 + \matelem{V}{kj} z_j) \approx 1+\matelem{V}{kj} z_j$, assuming a small variance, $\|\mat{V}\|_F \ll 1$. After ignoring the second-order and higher terms, (\ref{eq:rootn}) approximates
\begin{equation}
\label{eq:root_linear}
\sum_k w_k F_k + \sum_{k} w_k F_k (\matrow{\covsq}{k} \vecdot{z}+\matelem{V}{k1}z_1) = K,
\end{equation}
and $\partial d(\vecdot{z})/\partial z_j$ is obtained as the constant
\begin{equation}
\label{eq:sensitivity}
\frac{\partial\, d(\vecdot{z})}{\partial\, z_j}
\approx
\frac{\vect{\wf}^\transp \matcol{\covsq}{j}}{\vect{\wf}^\transp \matcol{\covsq}{1}}
= \frac{\vect{\wf}^\transp \mat{C}\, \matcol{Q}{j}}{  \vect{\wf}^\transp \mat{C}\, \matcol{Q}{1}}
\quad \text{for} \quad j\ge 2.
\end{equation}
Here, $\vect{\wf}$ is the normalized forward-adjusted weight vector,
$\wf_k \propto w_kF_k$, where $|\vect{\wf}|=1$.
The partial derivatives are minimized to zero when $\matcol{Q}{1}$ is aligned to the direction of $\mat{C}^\transp \vect{\wf}$ because the choice maximizes the denominator and makes the numerator zero from the orthogonality between $\matcol{Q}{1}$ and $\matcol{Q}{j}$ for $j\ge 2$. Therefore, the optimal first factor is determined:
\begin{equation}
\label{eq:V1}
\matcol{Q}{1} = \frac{\mat{C}^\transp\, \vect{\wf}}{\sqrt{\vect{\wf}\!^\transp \mat{\cov}\, \vect{\wf}}}
\quad \text{and} \quad
\matcol{V}{1} = \mat{C}\,\matcol{Q}{1} =
\frac{\mat{\cov}\, \vect{\wf}}{\sqrt{\vect{\wf}\!^\transp \mat{\cov}\, \vect{\wf}}}.
\end{equation}
This is equivalent to rotating the $z_1$ axis to the steepest ascending direction in the left-hand side of (\ref{eq:root_linear}), which is in agreement with the observation from \S~\ref{sec:example}. The other axes span the slowly varying dimensions, which are subject to costly numerical integrations. The overall computation cost can be minimized in this manner.

However, $\matcol{V}{1}$ in (\ref{eq:V1}) does not always conform to the earlier constraint $w_k \matelem{V}{k1}>0$. In the case of $w_k \matelem{V}{k1}\le 0$, for certain $k$, $\matelem{V}{k1}$ is adjusted by pushing it into the conforming region by
\begin{equation}
\label{eq:bound_adj}
\matelem{V}{k1}^\text{(adj)} = \begin{cases}
\mu \matelem{V}{k1} &\text{if}\quad w_k \matelem{V}{k1}>0 \\
\mu\, \varepsilon \,\text{sign}(w_k) \sqrt{\matelem{\cov}{kk}} &\text{if}\quad w_k \matelem{V}{k1}\le 0,
\end{cases}
\end{equation}
for a small $\varepsilon>0$ and rescaling factor $\mu>0$, thereby making $\matcol{Q}{1} = \mat{C}^{-1} \matcol{V}{1}^\text{(adj)}$ a unit vector ($\mu=1$ if no adjustment). Here, $\sqrt{\matelem{\cov}{kk}}$ is used as a characteristic scale of $\matelem{V}{k1}$ because $|\matelem{V}{k1}|\le\sqrt{\matelem{\cov}{kk}}$.

\subsection{Remaining factors}
The remaining columns $\matcol{\covsq}{j}$ for $j\ge 2$ are determined using singular value decomposition (SVD); this rearranges the columns orthogonally and in decreasing order of factor strength $|\matcol{\covsq}{j}|$.
This process is executed in the following two steps. First, an orthonormal rotation matrix is found, whose first column is the same as $\matcol{Q}{1}$. The computationally lightest choice is the Householder reflection matrix $\mat{R}$, which maps $\vect{e}_1 = (1,0,\cdots)^\transp$ to $\matcol{Q}{1}$ using the mirror image
\begin{equation}
\mat{R} = \mat{I} - 2\vect{v}\vect{v}^\transp \quad \text{where} \quad
\vect{v} = (\matcol{Q}{1}-\vect{e}_1)/|\matcol{Q}{1}-\vect{e}_1|.
\end{equation}
Thus, the first column of $\mat{C}\mat{R}$ is equal to $\matcol{V}{1}$.
Second, the remaining columns of $\mat{C}\mat{R}$ are rearranged via the reduced-size SVD,
$\mat{C} \, (\matcol{R}{2}\; \cdots\; \matcol{R}{N}) = \vecdot{U}\, \vecdot{D}\, \vecdot{Q}^\transp$,
where $\vecdot{D}$ is an $(N-1)\times(N-1)$ diagonal matrix with the (non-negative) singular values in decreasing order, and
$\vecdot{U}$ and $\vecdot{Q}$ are the $N\times(N-1)$ and $(N-1)\times(N-1)$ matrices, respectively, satisfying $\vecdot{U}^\transp\vecdot{U} = \vecdot{Q}^\transp\vecdot{Q} = \mat{I}$. Finally, the full $\mat{\covsq}$ is obtained by the following column-wise concatenation
\begin{equation}
\mat{\covsq} = (\matcol{V}{1}\;\; \vecdot{U}\vecdot{D} ),
\end{equation}
and the corresponding $\mat{Q}$ as
\begin{equation}
\mat{Q} = \mat{R}
\left(\begin{array}{c@{\;\;}c} 1 & \vect{0}^\transp \\ \vect{0} & \vecdot{Q} \end{array} \right),
\end{equation}
where $\vect{0}$ is the zero vector.
Because of the linearized assumption, the choice of $\mat{V}$ is independent of the strike price $K$; this ensures that if $\mat{V}$ is computed once, then it can be used for options with multiple strike prices.

\section{Remarks on Asian options \label{sec:asianopt}}
This section discusses a few implications of the method employed here in the context of Asian options.
First, since the covariance $\mat{\cov}$ is from the self-correlation in a Brownian motion, the Cholesky decomposition $\mat{C}$ is simply computed as
\begin{equation*}
\matelem{C}{kj} =
\begin{cases}
0 & \text{if}\quad k<j\\
\sigma \sqrt{\,t_j-t_{j-1}} & \text{if}\quad  k\ge j \quad(t_0=0).
\end{cases}
\end{equation*}
Therefore, there is no computational burden for a large $N$.

Second, all elements of $\matcol{\covsq}{1}$ in (\ref{eq:V1}) are positive because all elements of $\mat{\cov}$ and $\vect{\wf}$ are positive for Asian options.
Essentially, the selected $\matcol{\covsq}{1}$ is not compromised by the adjustment step of (\ref{eq:bound_adj}).

\begin{figure}
	\caption{First three terms of a standard Brownian motion series expansion.
	The solid (blue) lines are the factors $V_j(t)$ obtained using the method presented here, computed with $N=50$, while the dashed (red) lines are the factors $V^{(\text{KL})}_j(t)$ obtained from the Karhunen-–Lo\`eve expansions in (\ref{eq:kl_expansion}).
} \label{fig:asian_fact}
	\centering
	\includegraphics[width=0.48\textwidth]{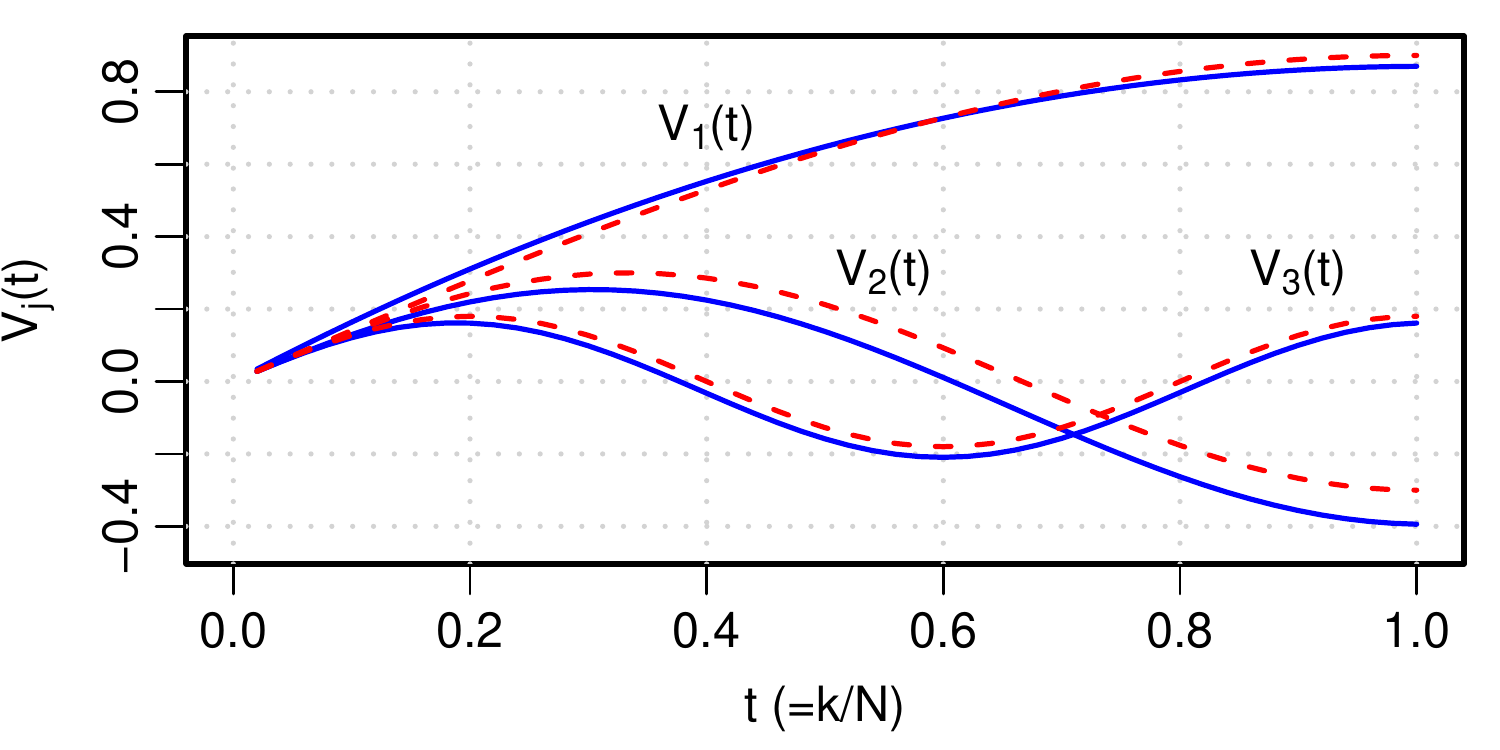}
\end{figure}

Third, the columns of factor matrix $\mat{\covsq}$ can be interpreted as a series representation of the Brownian motion $\sigma W(t)$ on the discretized time set $\{t_k\}$, $\sigma\,\big( W(t_1), \cdots, W(t_N)\big)^\transp \distequal \mat{\covsq} \vect{z}$.
If $V_j(t)$ is the continuum limit of $\matelem{V}{kj}$, as $N \rightarrow \infty$, with $t=k/N$ fixed for $\sigma=1$, then the set $\{V_j(t)\}$ would serve as a series expansion of $W(t)$:
\begin{equation}
W(t) \distequal \sum_{j=1}^\infty V_j(t)\; z_j,
\end{equation}
for independent standard normals $\{z_j\}$. Therefore, it is worth comparing this expansion to the well-known Karhunen--Lo\`eve expansions:
\begin{equation}
\label{eq:kl_expansion}
W(t) \distequal \sum_{j=1}^\infty V_j^\text{(KL)}(t)\;z_j \quad\text{for}\quad V_j^\text{(KL)}(t) = \frac{\sqrt{2}}{(j-\frac12)\,\pi}\sin\big((j-\frac12)\,\pi t\big).
\end{equation}
Figure~\ref{fig:asian_fact} shows the first three factors from the two expansions. The terms are similar, but with a subtle difference in the first factor. For the normalized parameters, that is, for $\sigma=1$, $t_k = k/N\; (T=1)$, $w_k=1/N$, and $r=q=0$, $\matelem{V}{1}(t)$ is given by the continuous version of (\ref{eq:V1}):
\begin{equation} \label{eq:V1fn}
\matelem{V}{1}(t) = \frac{\int_0^1 \min(t,u) du}{\sqrt{\int_0^1\int_0^1 \min(s,u) duds}} = \sqrt3 \Big(t-\frac12 t^2\Big).
\end{equation}
Since the Karhunen--Lo\`eve expansion is the PCA of $W(t)$ in the functional space, $V_1^{\text{(KL)}}(t)$ is chosen to maximize the $L^2$-norm, $\int_0^1 \matelem{V}{1}^2(t) dt$. Therefore, $\|V_1^\text{(KL)}(t)\|_2 = 2/\pi \,(\approx 0.6366)$ is larger than $\|\matelem{V}{1}(t)\|_2 = \sqrt{2/5} \,(\approx 0.6325)$.
On the other hand, $V_1(t)$ is chosen to maximize $\int_0^1 V_1(t) dt$.
Thus, $\int_0^1 V_1(t)\,dt = 1/\sqrt{3} \,(\approx 0.5774)$ is larger than $\int_0^1 V^{(\text{KL})}_1(t)\,dt = 4\sqrt{2}/\pi^2 \,(\approx 0.5732)$.
The continuous representation (\ref{eq:V1fn}) is valid only under constant weight, and no longer holds if, for example, $r\neq 0$ or $w_k$ is not constant. Therefore, generally $\matcol{V}{1}$ is numerically computed instead of using (\ref{eq:V1fn}).

Fourth, dimensionality reduction is critically effective in Asian options because the dimension $N$ is large. For a one-year maturity, the monthly averaging, weekly averaging, and daily averaging correspond approximately to $N=12$, $N=50$, and $N=250$, respectively. Even if a few nodes per dimension are used, the total number of nodes becomes prohibitively large.
However, the first several factors explain most of the variance in option prices, while the remaining factors have negligible impact. Table~\ref{tab:asian_var} shows the portion of the explained variance, $\sum_{j=1}^{N'} |\matcol{V}{j}|^2/\| \mat{V} \|_F^2$, as a function of increasing $N'$. The first factor, $\matcol{V}{1}$, accounts for the largest part, about 80\%, with the cumulative portion reaching about 96\% at $N'=5$. 
The numerical results in \S~\ref{sec:result} show that five dimensions, one under analytic and four under numerical integration, are sufficient for accurate option pricing.
If the averaging feature is meant to avoid market manipulation, as is often the case, averaging should start sometime near maturity. Since the correlation in this case is overall higher compared to that in the case of immediate averaging, the explained variance portion is higher and the dimensionality reduction more effective.

Finally, the valuation of continuously monitored Asian options is cast into discrete monitoring. For a better integration convergence over time, the Simpson's rule weights are used rather than constant weights.
For the discretization step $\Delta T$ such that $N = T/\Delta T$ is even, the observation time and weights are given as
\begin{equation}
\label{eq:simpson}
t_k = k\Delta T, \quad
w_k = \begin{cases}
4 \Delta T/(3T) & \text{if $k$ is odd} \\
2 \Delta T/(3T) & \text{if $k$ is even}
\end{cases}
\quad \text{for}\quad 0\le k\le N,
\end{equation}
with the exception of the weights at the two end points, $w_0 = w_N = \Delta T / (3T)$.

\begin{table}
	\caption{Ratios of explained variance in percentage (\%) as function of first $N'$ dimensions. It is assumed that $r=q=0$ and $w_k=1/N$.}
	\label{tab:asian_var}
	\begin{center}
		\begin{tabular}{c|c|c|c|c|c} \hline
			$N \;\backslash\; N'$ & 1 & 2 & 3 & 4 & 5 \\ \hline
			12 & 80 & 90 & 94 & 96 & 97 \\
			50 & 80 & 90 & 93 & 95 & 96 \\
			250 & 80 & 90 & 93 & 95 & 96 \\ \hline
		\end{tabular}
	\end{center}
\end{table}

\section{Numerical results \label{sec:result}}

\begin{table}
	\caption{Parameter sets for test cases. The parameter noted by $\ast$ is to be varied in the test; the value superscripted by $\ast$ represents the base value when not varied.}
	\begin{center}
		\label{tab:test_sets}
		\begin{tabular}{c||c|c|c|c|c|c|c|c|c} \hline
			\multirow{2}{*}{Label} & $N$ & $t_k$ or $T$ & $S_k(0)$ & $K$ & $w_k$ & $\sigma_k\, (\%)$ & $\rho_{k\neq j}\, (\%)$ & $q_k\, (\%)$ & $r\, (\%)$ \\ \cline{2-10}
			& \multicolumn{9}{l}{References that test same parameter set} \\	\hline \hline
			\multirow{2}{*}{\textbf{S1}} & 2 & 1 & $(100, 96)$ & $\ast$ & $(1,-1)$ & $(20,10)$& 50 & 5 & 10 \\ \cline{2-10}
			& \multicolumn{9}{l}{\citet{dempster2002spread,hurdzhou2010spread,caldana2013spread}} \\ \hline \hline
			\multirow{2}{*}{\textbf{S2}} & 2 & 1 & $(200, 100)$ & $100$ & $(1,-1)$ & $(15,30)$& $\ast$ & 0 & 0 \\ \cline{2-10}
			& \multicolumn{9}{l}{No previous reference} \\ \hline \hline
			\multirow{2}{*}{\textbf{B1}} & 4 & 5 & 100 & $100^\ast$ & $1/4$ & $40^\ast$ & $50^\ast$ & 0 & 0 \\ \cline{2-10}
			& \multicolumn{9}{l}{ \citet{wilmott_basket,caldana2016basket} } \\ \hline \hline
			\multirow{2}{*}{\textbf{B2}} & 7 & $\ast$ & 100 & $\ast$ & \multicolumn{4}{c|}{ Table~\ref{tab:G7_set}} & 6.3 \\ \cline{2-10}
			& \multicolumn{9}{l}{ \citet{milevsky1998basket,zhouwang2008} } \\ \hline \hline
			\multirow{2}{*}{\textbf{A1}} & 50 & $k/50\;(k\ge 0)$ & 100 & $\ast$ & 1/51 & $\ast$ & $\cdot$ & 0 & 10 \\ \cline{2-10}
			& \multicolumn{9}{l}{ \citet{levy1992average,benhamou2000fast,cerny2011asian}  } \\
			\hline \hline
			\multirow{3}{*}{\textbf{A2}} & $\ast$ & $k/N\;(k\ge 0)$ & 100 & $\ast$ & 1/(N+1) & 17.801 & $\cdot$ & 0 & 3.67 \\ \cline{2-10}
			& \multicolumn{9}{l}{ \citet{fusai2008pricing,cerny2011asian,fusai2011pricing}, } \\
			& \multicolumn{9}{l}{ \citet{cai2013asian,zhang2013efficient} } \\  \hline \hline
		\end{tabular}
	\end{center}
\end{table}

\begin{table}
	\caption{Remaining parameters for \textbf{B2}}
	\label{tab:G7_set}
	\begin{center}
		\begin{tabular}{c|c|rrrrrr|c} \hline
			$w_k$ & $\sigma_k (\%)$ & \multicolumn{6}{c|}{$\rho_{kj}(\%)$ for $k<j$ } & $q_k (\%)$ \\ \hline
			0.10 & 11.55 & \hspace{5mm} 35 & 10 & 27 & 4 & 17 & 71 & 1.69\\
			0.15 & 20.68 &  & 39 & 27 & 50 & -8 & 15 & 2.39\\
			0.15 & 14.53 &  &  & 53 & 70 & -23 & 9 & 1.36\\
			0.05 & 17.99 &  &  &  & 46 & -22 & 32 & 1.92\\
			0.20 & 15.59 &  &  &  &  & -29 & 13 & 0.81\\
			0.10 & 14.62 &  &  &  &  &  & -3 & 3.62\\
			0.25 & 15.68 &  &  &  &  &  &  & 1.66\\
			\hline
		\end{tabular}
	\end{center}
\end{table}	

The method is implemented in R (Ver. 3.3.2, 64-bit) on a personal computer running the Windows 10 operating system with an Intel core i5 2.2 GHz CPU and 8 GB RAM. Seven parameter sets are tested, the first six of which are described in Tables~\ref{tab:test_sets} and \ref{tab:G7_set}. The sets are labeled as \textbf{S} for spread, \textbf{B} for basket, and \textbf{A} for Asian options.
The set \textbf{A3} for continuously monitored Asian options is separately displayed in Table~\ref{tab:asian3} along with the results.
Except for set \textbf{S2}, the parameters are the same ones used in previous studies to enable easier comparison.

The numerical results are reported in Tables~\ref{tab:spread1}$\sim$\ref{tab:asian3}. Except for the three Asian option sets, two versions of the prices are computed. One version is the ``fast'' price, for which the minimal GHQ nodes are used with a target precision of 3$\sim$4 decimals, which is both practical and sufficient. The computational cost for the fast price is inexpensive. The other version is the converged price; the fast price error is measured against this. The converged price is obtained as the node sizes are increased. Seven decimal precisions are targeted for the converged price.
For the Asian option cases (Tables~\ref{tab:asian1}$\sim$\ref{tab:asian3}), only the fast price is reported; the error is measured from the previous studies: \citet{cerny2011asian} for the discrete cases (\textbf{A1}, and \textbf{A2}) and \citet{linetsky2004spectral} for the continuous case (\textbf{A3}). All prices in the tables are the present values of the call options; that is, the forward value discounted by $e^{-rT}$.

Some tables show the used factor matrix $\mat{\covsq}$ in the following representation to provide extra properties besides the matrix itself.
\begin{equation} \label{eq:V_display}
\textstyle
\begin{array}{c@{\;}|@{\;\;}cccc@{\;\;}|@{\;\;}c}
\vect{\wf}^\transp \matcol{V}{1} & |\matcol{V}{1}| & \multicolumn{2}{c}{\cdots} & |\matcol{V}{N}| & \| \mat{\covsq} \|_F\,=(\sum_k \matelem{\cov}{kk})^{1/2} \\ \hline
\wf_1 & \matelem{\covsq}{11} & \multicolumn{2}{c}{\cdots} & \matelem{\covsq}{1N} & |\matrow{V}{1}|\,=\sqrt{\cov_{11}} \\
\vdots & \vdots & \multicolumn{2}{c}{\matelem{\covsq}{kj}} & \vdots & \vdots  \\
\wf_N & \matelem{\covsq}{N1} & \multicolumn{2}{c}{\cdots} & \matelem{\covsq}{NN} & |\matrow{V}{N}|\,=\sqrt{\cov_{NN}} \\ \hline
& \cdot & M_2 & \cdots & M_N & M = \prod_{j\ge 2} M_j.
\end{array}
\end{equation}
The center itself is $\mat{\covsq}$. The upper and right-hand side panels show the $L^2$-norm of the columns and rows, respectively, while the upper right corner is the Frobenius norm. The left-hand side panel shows the forward-adjusted weight vector $\vect{\wf}$, and the upper left corner shows the dot product $\vect{\wf}^\transp \matcol{V}{1}$. The lower panel shows the node size $M_j$ for the $j$-th dimension of the $j\ge 2$, with the total size $M$ in the lower right corner.

For implementation, $M_j$ must be chosen in an economic manner. The node sizes need not be the same for all dimensions. Therefore, $M_j$ can be selected to ensure that all dimensions homogeneously achieve a similar level of accuracy. A general guideline is that the density of the nodes should be proportional to $|\partial d(\vecdot{z})/\partial z_j|$ in order to effectively capture the change in $d(\vecdot{z})$. To this extent, the following rule is used to systematically determine $M_j$ in the numerical tests:
\begin{equation}
\label{eq:rule_M}
M_j = \Big[ \frac{ |\matcol{\covsq}{j}|}{| \vect{\wf}^\transp \matcol{\covsq}{1}|} \lambda + 1\Big]
\quad \text{for} \quad j \ge 2,
\end{equation}
where $[x]$ is the nearest integer of $x$ and $\lambda$ is the coefficient for the level of accuracy. The ratio $|\matcol{\covsq}{j}|/|\vect{\wf}^\transp \matcol{\covsq}{1}|$  is obtained by applying the Cauchy-Schwarz inequality,
$|\vect{\wf}^\transp\matcol{V}{j}|<|\vect{\wf}|\cdot|\matcol{V}{j}|=|\matcol{V}{j}|$,
to (\ref{eq:sensitivity}), which is thereby understood as an approximate upper bound of $|\partial d(\vecdot{z})/\partial z_j|$. This rule serves also as a criterion for dimension reduction: if $M_j=1$ for some $j$, the dimension can be truncated according to \S~\ref{sec:reduction}.
Moreover, this rule is independent of $K$; it ensures that if $\{\vecdot{z}_m\}$ and $\{w_m\}$ are computed once, they can be used for options with multiple values of $K$.
For Asian options, however, $M_j=3$ for $2\le j\le 5$ ($M=81$) is empirically found to work very well; thus, the test cases for Asian options do not resort to (\ref{eq:rule_M}).

\subsection{Spread options}
In \textbf{S1}, the spread call options are priced for varying strikes from $K=0$ (in-the-money) to $4$ (at-the-money). For the speed of convergence, the node size is increased for the second dimension from $M_2=2$. As shown in Table~\ref{tab:spread1}, convergence is extremely fast. While the prices with $M_2=2$ are already accurate,
they converge within seven decimals at $M_2=3$. The table also shows that the control variate correction of \S~\ref{sec:CV} can further reduce error.
While \citet{hurdzhou2010spread} and \citet{caldana2013spread} show accuracy similar to the $M_2=3$ result, their methods have to evaluate the expensive Fourier inversion in two and one dimensions, respectively.
The risk factor matrix is presented in Table~\ref{tab:spread1}(b). In this example, the adjustment step (\ref{eq:bound_adj}) is triggered for $\matelem{V}{21}$ with $\varepsilon = 0.01$.
In addition, this study independently implements the analytic approximation methods of \citet{bjerksund2014spread}, \citet{lo2015pricing}, and \citet{li2008spread} for spread options.\footnote{For \citet{lo2015pricing}, the ``Strang's splitting approximation I'' method in the reference is implemented; this is given in an explicit formula.} All the three methods work well on \textbf{S1}, showing errors in the order of $10^{-6}$ or less.

In \textbf{S2}, the at-the-money spread call options are priced for varying correlations $\rho_{12}$ from $90\%$ to $-90\%$. The results are shown in Table~\ref{tab:spread2}(a). This parameter set is designed to make the first asset to follow a displaced GBM, $dS_1(t) = \sigma_{1} (S_1(t)+L)\,dW_1(t)$ with $L=100$ to demonstrate a negative implied volatility skew. Therefore, the true parameters are similar to those of $S_2$, as $S_1(0)=100$ and $\sigma_1 \approx 30\%$, and $K=100$ corresponds to the at-the-money strike price.
The factor matrix $\mat{\covsq}$ changes with the change in $\rho_{12}$ and $M_2$ is determined according to (\ref{eq:rule_M}). For a fast price, $M_2$ varies from $17$ to $2$ when $\lambda=3$ is used. Since the components in $\matcol{V}{1}$ must have opposite signs for spread options, its norm weakens under positive correlation, thus requiring denser nodes. The quadrature size rule (\ref{eq:rule_M}) works reasonably well, although it tends to over-allocate nodes under high correlation.
The price quickly converges as $\lambda$ is increased, to achieve the seven-decimal precision at $\lambda=9$. Tables~\ref{tab:spread2}(b) and (c) show $\mat{\covsq}$ for $\rho_{12} = 90\%$ and $-90\%$, respectively. Note that the two columns switch places with a sign change.
The performance of analytic approximation methods in \textbf{S2} is not as good as that in \textbf{S1}; this might be attributed to the displaced GBM features, that is, the significant difference between two spot prices and the big strike price. Although the method of \citet{li2008spread} is much more accurate than the other two methods, it illustrates the limitation of analytic approximation.

\begin{table}
	\caption{Numerical results for \textbf{S1} are as follows: (a) converged prices (CPs) and fast price errors (FP Err) with and without a control variate for varying $K$, and (b) factor matrix $\mat{\covsq}$ (see (\ref{eq:V_display}) for representation).}
	\label{tab:spread1}
	\begin{center}
		(a) \;
		\begin{tabular}[t]{c||c||c|c||c|c} \hline
			\multirow{2}{*}{$K$}& CP & \multicolumn{2}{c||}{FP Err w/ CV} & \multicolumn{2}{c}{FP Err w/o CV}\\ \cline{2-6}
			& $M_2=4$ & $M_2=3$ & $M_2=2$ & $M_2=3$ & $M_2=2$ \\ \hline
			0.0 & 8.5132252 & $-$7.4e-9 & $-$3.0e-6 & $-$1.3e-7 & $-$1.1e-4 \\
			0.4 & 8.3124607 & $-$8.2e-9 & $-$3.5e-6 & $-$1.3e-7 & $-$1.1e-4 \\
			0.8 & 8.1149938 & $-$9.0e-9 & $-$4.0e-6 & $-$1.3e-7 & $-$1.1e-4 \\
			1.2 & 7.9208198 & $-$9.8e-9 & $-$4.7e-6 & $-$1.3e-7 & $-$1.1e-4 \\
			1.6 & 7.7299325 & $-$1.1e-8 & $-$5.3e-6 & $-$1.3e-7 & $-$1.1e-4 \\
			2.0 & 7.5423239 & $-$1.1e-8 & $-$6.0e-6 & $-$1.3e-7 & $-$1.1e-4 \\
			2.4 & 7.3579843 & $-$1.2e-8 & $-$6.7e-6 & $-$1.3e-7 & $-$1.1e-4 \\
			2.8 & 7.1769024 & $-$1.2e-8 & $-$7.5e-6 & $-$1.3e-7 & $-$1.1e-4 \\
			3.2 & 6.9990651 & $-$1.3e-8 & $-$8.2e-6 & $-$1.3e-7 & $-$1.1e-4 \\
			3.6 & 6.8244581 & $-$1.3e-8 & $-$9.0e-6 & $-$1.3e-7 & $-$1.1e-4 \\
			4.0 & 6.6530651 & $-$1.3e-8 & $-$9.7e-6 & $-$1.3e-7 & $-$1.1e-4 \\
			\hline
		\end{tabular} \\ \vspace{1em}
		
		(b)
		\begin{tabular}[t]{r|rc|c}
			0.125 & 0.172 & 0.143 & 0.224 \\ \hline
			0.721 & 0.172 & 0.102 & 0.200 \\
			$-$0.693 & $-$0.001 & 0.100 & 0.100 \\ \hline
			& \multicolumn{1}{c}{$\cdot$} & 4 & 4
		\end{tabular} \\ \vspace{1ex}
		
	\end{center}
\end{table}

\begin{table}
	\caption{Numerical results for \textbf{S2} with varying correlation $\rho_{12}$. The converged prices (CPs) and fast price errors (FP Err) are shown in (a). Fast prices are obtained using $\lambda=3$ and converged price are obtained using $\lambda=9$. The errors of the three approximation methods are also provided for reference: \citet{bjerksund2014spread} (BjSt Err), \citet{lo2015pricing} (Lo Err), and \citet{li2008spread} (LDZ Err).
	Factor matrices $\mat{\covsq}$ for $\rho_{12}=90\%$ and $-90\%$ are displayed in (b) and (c), respectively.}
	\label{tab:spread2}
	\begin{center}
		(a)\;
		\begin{tabular}[t]{r||r|c|r||c|c|c} \hline
			$\rho_{12}\,(\%)$ & \multicolumn{1}{c|}{CP} & \multicolumn{1}{c|}{FP Err} & $M_2$ & BjSt Err & Lo Err & LDZ Err \\ \hline
			90 & 5.4792720 & $-$1.5e-8 & 17 & $-$1.3e-1 & $+$1.4e-2 & $-$1.9e-2 \\
			70 & 9.3209439 & $+$3.7e-8 & 10 & $-$6.3e-2 & $+$1.0e-2 & $-$1.0e-4 \\
			50 & 11.9804918 & $+$2.2e-7 & 7 & $-$4.0e-2 & $+$1.0e-2 & $+$7.9e-4 \\
			30 & 14.1425869 & $-$4.0e-7 & 6 & $-$2.8e-2 & $+$8.8e-3 & $+$7.7e-4 \\
			10 & 16.0102190 & $-$1.4e-7 & 5 & $-$2.0e-2 & $+$6.6e-3 & $+$6.1e-4 \\
			$-$10 & 17.6770249 & $+$5.2e-6 & 4 & $-$1.6e-2 & $+$3.7e-3 & $+$4.5e-4 \\
			$-$30 & 19.1954201 & $+$1.5e-6 & 4 & $-$1.4e-2 & $-$3.6e-5 & $+$3.1e-4 \\
			$-$50 & 20.5982705 & $-$8.0e-6 & 3 & $-$1.3e-2 & $-$4.3e-3 & $+$1.9e-4 \\
			$-$70 & 21.9077989 & $-$1.7e-6 & 3 & $-$1.4e-2 & $-$9.2e-3 & $+$1.0e-4 \\
			$-$90 & 23.1398674 & $-$8.3e-5 & 2 & $-$1.5e-2 & $-$1.5e-2 & $+$2.0e-5 \\
			\hline
		\end{tabular}\\ \vspace{1em}
		
		(b)\;
		\begin{tabular}[t]{r|rc|c}
			0.060 & 0.075 & 0.327 & 0.335 \\ \hline
			0.894 & 0.034 & 0.146 & 0.150 \\
			$-$0.447 & $-$0.067 & 0.292 & 0.300 \\ \hline
			& \multicolumn{1}{c}{$\cdot$} & 17 & 17
		\end{tabular} \hspace{1ex}
		(c)\;
		\begin{tabular}[t]{r|rc|c}
			0.262 & 0.327 & 0.075 & 0.335 \\ \hline
			0.894 & 0.146 & 0.034 & 0.150 \\
			$-$0.447 & $-$0.292 & 0.067 & 0.300 \\ \hline
			& \multicolumn{1}{c}{$\cdot$} & 2 & 2
		\end{tabular}
	\end{center}
\end{table}

\subsection{Basket Option}
Parameter sets \textbf{B1} and \textbf{B2} have been frequently used as benchmarks in previous studies. Here, the convergent option values are reported for the first time.
Parameter set \textbf{B1} is extracted from \citet{wilmott_basket}, where the performance results of several analytic approximation methods are compared.
Tables~\ref{tab:basket_K} and \ref{tab:basket_corr} provide the results for varying $K$ and $\rho_{k\neq j}$, respectively, from the base parameter values. Moreover, Table~\ref{tab:basket_vol1} tests the inhomogeneous volatilities by varying $\sigma_k$ simultaneously for $1\le k\le 3$, while keeping $\sigma_4 = 100\%$ fixed.
In all tests, $\lambda=9$ give consistent fast prices, which are more accurate than the benchmark Monte Carlo prices of \citet{wilmott_basket}; moreover, the computation cost is much cheaper. The test in Table~\ref{tab:basket_K} shows that $\lambda=9$ corresponds to $M_j=5$ for $2\le j\le 4$ ($M=125$). In addition, the node sizes in Tables~\ref{tab:basket_corr} and \ref{tab:basket_vol1} are shown in their respective last columns. Contrary to the spread option, a lower correlation requires denser nodes in a basket option.\footnote{In the test, the lower bound of the correlation $\rho_{k\neq j}$ is $-1/3$; this is to ensure that the covariance matrix is positive-semidefinite.} Also note that only a few nodes per dimension produce accurate prices in the inhomogeneous volatility test. No approximation method produces satisfactory precision in the same test~\citep{wilmott_basket}.

The parameter set \textbf{B2}, referred to as the G-7 indices basket option, has been tested in \citet{milevsky1998basket} and \citet{zhouwang2008}.
The call option prices for varying $K$ and $T$ are shown in Table~\ref{tab:basket_G7}(a) and the factor matrix $\mat{\covsq}$ is shown in Table~\ref{tab:basket_G7}(b). The current method works well for this $N=7$ case. The error of the fast prices with $\lambda=3$ ($M=432$) is in the order of $10^{-4}$ at most, which is smaller than the standard error of the Monte Carlo simulation in \citet{zhouwang2008}. The seven-digit convergent prices are computed with $\lambda=12$ ($M\approx 1.2\times 10^5$).

\begin{table}
	\caption{Numerical results for \textbf{B1} with varying $K$ are as follows: (a) converged prices (CPs) and fast price errors (FP Err), and (b) factor matrix $\mat{\covsq}$. Fast prices are obtained using $\lambda=9$, which results in $M_{j\ge 2}=5$ ($M=125$).}
	\label{tab:basket_K}
	\begin{center}
		(a)
		\begin{tabular}[t]{c||c|c} \hline
			$K$ & CP & FP Err \\ \hline
			50 & 54.3101761 & $-$2.0e-4 \\
			60 & 47.4811265 & $-$2.5e-4 \\
			70 & 41.5225192 & $-$2.6e-4 \\
			80 & 36.3517843 & $-$2.4e-4 \\
			90 & 31.8768032 & $-$2.0e-4 \\
			100 & 28.0073695 & $-$1.3e-4 \\
			110 & 24.6605295 & $-$6.4e-5 \\
			120 & 21.7625789 & $+$8.4e-6 \\
			130 & 19.2493294 & $+$7.9e-5 \\
			140 & 17.0655420 & $+$1.4e-4 \\
			150 & 15.1640103 & $+$2.0e-4 \\
			\hline
		\end{tabular}
		(b)
		\begin{tabular}[t]{c|crrr|r}
			1.414 & 1.414 & 0.632 & 0.632 & 0.632 & 0.000 \\ \hline
			0.500 & 0.707 & $-$0.000 & $-$0.420 & 0.352 & 0.894 \\
			0.500 & 0.707 & $-$0.000 & $-$0.191 & $-$0.513 & 0.894 \\
			0.500 & 0.707 & $-$0.447 & 0.306 & 0.081 & 0.894 \\
			0.500 & 0.707 & 0.447 & 0.306 & 0.081 & 0.894 \\ \hline
			& $\cdot$ & \multicolumn{1}{c}{5} & \multicolumn{1}{c}{5} & \multicolumn{1}{c}{5} & \multicolumn{1}{|c}{125} \\
		\end{tabular} \\ \vspace{1ex}
		
	\end{center}
\end{table}

\begin{table}
	\caption{Numerical results for \textbf{B1} with simultaneously varying correlation $\rho_{k\neq j}$. The converged prices (CPs), fast price error (FP Err), and quadrature sizes are shown.
	Fast prices are obtained using $\lambda=9$.}
	\label{tab:basket_corr}
	\begin{center}
		\begin{tabular}{r||c|c|c} \hline
			$\rho_{k\neq j}\,(\%)$ & CP & FP Err & $M_{j\ge 2} (M)$ \\ \hline
			$-$10 & 17.7569163 & $-$4.9e-8 & 12, 12, 12 (1728) \\
			10 & 21.6920965 & $-$7.3e-6 & 7, 7, 7 (343) \\
			30 & 25.0292992 & $+$1.3e-4 & 6, 6, 6 (216) \\
			50 & 28.0073695 & $-$1.2e-4 & 5, 5, 5 (125) \\
			80 & 32.0412265 & $-$4.0e-4 & 3, 3, 3 (27) \\
			95 & 33.9186874 & $-$3.1e-3 & 2, 2, 2 (8) \\
			\hline
		\end{tabular}	
	\end{center}
\end{table}

\begin{table}
	\caption{Numerical results for \textbf{B1} having varying volatilities $\sigma_{k\le 3}$ with $\sigma_4=100\%$ fixed are as follows:
	(a) converged prices (CPs), fast price errors (FP Err), and node sizes, and (b) factor matrix $\mat{\covsq}$ for the case of $\sigma_{k\le 3}=10\%$. Fast prices are obtained using $\lambda=9$.}
	\label{tab:basket_vol1}
	\begin{center}
		(a) \;
		\begin{tabular}[t]{c||c|c|c} \hline
			\multicolumn{1}{c||}{$\sigma_{k\le 3}\,(\%)$}  & CP & FP Err & $M_{j\ge 2} (M)$ \\ \hline
			5 & 19.4590950 & $-$4.3e-4 & 3, 2, 2 (12) \\
			10 & 20.9682321 & $+$8.4e-4 & 4, 2, 2 (16) \\
			20 & 25.3794239 & $+$6.9e-4 & 5, 3, 3 (45) \\
			40 & 36.0485407 & $+$1.6e-3 & 6, 4, 4 (96) \\
			60 & 46.8189186 & $+$6.5e-3 & 6, 4, 4 (96) \\
			80 & 56.7772198 & $-$9.2e-3 & 5, 5, 5 (125) \\
			100 & 65.4256003 & $+$1.8e-4 & 5, 5, 5 (125) \\
			\hline
		\end{tabular}\\ \vspace{1em}
		(b) \;
		\begin{tabular}[t]{c|crrr|r}
			1.304 &	2.217 &	0.429 &	0.158 &	0.158 &	2.269 \\ \hline
			0.500 &	0.134 &	$-$0.124 &	0.129 &	$-$0.011 &	0.224 \\
			0.500 &	0.134 &	$-$0.124 &	$-$0.054 &	0.117 &	0.224 \\
			0.500 &	0.134 &	$-$0.124 &	$-$0.074 &	$-$0.106 &	0.224 \\
			0.500 &	2.205 &	0.371 &	$-$0.000 &	$-$0.000 &	2.236 \\ \hline
			& $\cdot$ & \multicolumn{1}{c}{4} & \multicolumn{1}{c}{2} & \multicolumn{1}{c}{2} & \multicolumn{1}{|c}{16} \\
		\end{tabular}
	\end{center}
\end{table}

\begin{table}
	\caption{Numerical results for \textbf{B2} with varying $K$ and $T$ are as follows: (a) converged prices (CPs) and fast price errors (FP Err), and (b) factor matrix $\mat{\covsq}$ for the $T=1$ case.}
	\label{tab:basket_G7}
	\begin{center}
		(a)
		\begin{tabular}[t]{c||r|r|r|r||c|c|c|c} \hline
			& \multicolumn{4}{c||}{CP ($\lambda=12$)} & \multicolumn{4}{c}{FP Err ($\lambda=3$)} \\ \cline{2-9}
			& \multicolumn{4}{c||}{$M_{j\ge 2} = (11, 10, 8, 7, 5, 4),\; M = 123200$} &
			\multicolumn{4}{c}{$M_{j\ge 2} = (4,3,3,3,2,2),\;M = 432$} \\ \hline
			$K$ & \multicolumn{1}{c|}{$T=0.5$} & \multicolumn{1}{c|}{$T=1$} & \multicolumn{1}{c|}{$T=2$} & \multicolumn{1}{c||}{$T=3$} & $0.5$ & $1$ & $2$ & $3$ \\ \hline
			80 & 21.6022546 & 23.1411627 & 26.0424328 & 28.6992602 & $-$1.6e-8 & $-$9.0e-7 & $-$1.0e-5 & $-$2.8e-5\\
			100 & 3.8828353 & 6.2216810 & 10.2156012 & 13.7425580 & $-$4.3e-5 & $-$1.1e-4 & $-$2.5e-4 & $-$3.7e-4\\
			120 & 0.0235189 & 0.3535584 & 2.0570044 & 4.4578389 & $-$5.8e-6 & $-$7.9e-5 & $-$4.1e-4 & $-$7.8e-4\\
			\hline
		\end{tabular} \\ \vspace{1em}	
		
		(b)
		\begin{tabular}[t]{c|rrrrrrr|r}
			0.23 & 0.26 & 0.20 & 0.17 & 0.14 & 0.12 & 0.08 & 0.06 & 0.42 \\ \hline
			0.24 & 0.07 & -0.06 & -0.01 & -0.03 & 0.02 & -0.03 & 0.05 & 0.12 \\
			0.36 & 0.14 & 0.05 & 0.12 & -0.05 & 0.05 & 0.00 & -0.01 & 0.21 \\
			0.36 & 0.09 & 0.08 & -0.02 & 0.04 & -0.02 & -0.06 & -0.01 & 0.15 \\
			0.12 & 0.10 & 0.06 & -0.10 & 0.00 & 0.09 & 0.01 & 0.00 & 0.18 \\
			0.49 & 0.11 & 0.09 & 0.00 & 0.02 & -0.05 & 0.04 & 0.02 & 0.16 \\
			0.24 & 0.00 & -0.09 & 0.05 & 0.10 & 0.03 & 0.01 & 0.00 & 0.15 \\
			0.61 & 0.10 & -0.09 & -0.06 & -0.05 & -0.02 & 0.01 & -0.02 & 0.16 \\
			\hline
			& \multicolumn{1}{c}{$\cdot$} & \multicolumn{1}{c}{4} & \multicolumn{1}{c}{3} & \multicolumn{1}{c}{3}
			& \multicolumn{1}{c}{3} & \multicolumn{1}{c}{2}	& \multicolumn{1}{c|}{2} & \multicolumn{1}{c}{432}
		\end{tabular}
		
	\end{center}
\end{table}

\subsection{Asian options}
Three parameter sets for the Asian options are tested. The results for the discrete monitoring sets \textbf{A1} and \textbf{A2} are reported in Tables~\ref{tab:asian1} and \ref{tab:asian2}, respectively.\footnote{A parameter set, which is almost the same as \textbf{A1}, except for $r=4\%$, has been popularly tested in the literature as well; for example, \citet{cerny2011asian,fusai2011pricing} and \citet{cai2013asian}. They should not be confused.}
In Table~\ref{tab:asian1}, $K$ and $\sigma$ are varied, whereas in Table~\ref{tab:asian2}, $K$ and $N$ are varied. The errors are measured from \citet{cerny2011asian}, where the results are reported with an accuracy of $10^{-7}$. The fast prices computed with only 81 nodes over five dimensions ($M_j=3$ for $2\le j \le 5$) have errors in the order of $10^{-4}$ or less. The overall underpricing is due to a decline in variance from dimensionality reduction.
In \textbf{A1} and \textbf{A2}, the average computation time per option price is $0.01$, $0.02$, and $0.06$ seconds for $N=12,\; 50$, and $250$, respectively. This can be accelerated when the prices for multiple values of $K$ are computed together because the computations for $\mat{\covsq}$, $\{\vecdot{z}_m\}$, and $\{h_m\}$ are not repeated. Although a direct comparison on CPU time is difficult to obtain because of difference in computing environments, \citet{cerny2011asian} report 1 to 0.3 seconds as $\sigma$ varies from $10\%$ to $50\%$ for computing $N=50$ cases with a five-decimal precision.
Table~\ref{tab:asian3} reports the result for the continuous monitoring set \textbf{A3}. Because time is discretized with $\Delta T = 1/200$, $N=200$ for $T=1$ and $N=400$ for $T=2$. The fast prices computed with 81 nodes match the values of \citet{linetsky2004spectral}, computed with up to 10-decimal accuracy.
The computation time per option is $0.04$ and $0.18$ seconds for $N=200$ and $400$ cases, respectively.

\begin{table}
	\caption{Numerical results for \textbf{A1} with varying strike price $K$ and volatility $\sigma$. The fast prices (FPs) with $M_{2\le j\le 5}=3$ and errors (Err) are shown. Errors are measured from \citet{cerny2011asian}, where the values are reported with seven-decimal precision.}
	\label{tab:asian1}
	\begin{center}
		\begin{tabular}{c||r|r|r||r|r|r} \hline
			\multirow{2}{*}{$K$} & \multicolumn{3}{c||}{FP} & \multicolumn{3}{c}{Err $(\times 10^{-7})$} \\ \cline{2-7}
			& \multicolumn{1}{c|}{$\sigma = 10\%$} & \multicolumn{1}{c|}{$\sigma = 30\%$} & \multicolumn{1}{c||}{$\sigma = 50\%$} & $10\%$ & $30\%$ & $50\%$ \\ \hline
			80 & 22.7771749 & 23.0914273 & 24.8242382 & 0 & $-$105 & $-$199\\
			90 & 13.7337771 & 15.2207525 & 18.3316585 & $-$2 & $-$85 & $-$155\\
			100 & 5.2489922 & 9.0271796 & 13.1580058 & $-$5 & $-$92 & $-$398\\
			110 & 0.7238317 & 4.8348903 & 9.2344356 & $-$7 & $-$168 & $-$778\\
			120 & 0.0264089 & 2.3682616 & 6.3718411 & $-$3 & $-$238 & $-$1125\\
			\hline
		\end{tabular}
	\end{center}
\end{table}

\begin{table}
	\caption{Numerical results for \textbf{A2} for varying strike price $K$ and number of observations $N$. Fast prices (FPs) with $M_{2\le j\le 5}=3$ and errors (Err) are shown. Errors are measured from \citet{cerny2011asian}, where the values are reported with seven-decimal precision.}
	\label{tab:asian2}
	\begin{center}
		\begin{tabular}{c||r|r|r||r|r|r} \hline
			\multirow{2}{*}{$K$} & \multicolumn{3}{c||}{FP} & \multicolumn{3}{c}{Err $(\times 10^{-7})$} \\ \cline{2-7}
			& \multicolumn{1}{c|}{$N=12$} & \multicolumn{1}{c|}{$N=50$} & \multicolumn{1}{c||}{$N=250$} & \multicolumn{1}{c|}{$12$} & \multicolumn{1}{c|}{$50$} & \multicolumn{1}{c}{$250$} \\ \hline
			90 & 11.9049132 & 11.9329355 & 11.9405604 & $-$25 & $-$27 & $-$28\\
			100 & 4.8819577 & 4.9372004 & 4.9521546 & $-$39 & $-$24 & $-$23\\
			110 & 1.3630326 & 1.4025110 & 1.4133626 & $-$54 & $-$45 & $-$44\\
			\hline
		\end{tabular}
	\end{center}
\end{table}

\begin{table}
	\caption{Parameters and numerical results for \textbf{A3}. For all the seven cases, $K=2$ and $q=0$.
	Continuous monitoring is discretized by Simpson's rule (\ref{eq:simpson}) with $\Delta T = 1/200$.
	Fast prices (FPs) are computed with $M_{2\le j\le 5}=3$ and errors (Err) are measured from \citet{linetsky2004spectral}, where the values are reported with 10-decimal precision.}
	\label{tab:asian3}
	\begin{center}
		\begin{tabular}{c||c|c|c|c||c|r} \hline
			Case & $T$ & $S_0$ & $\sigma (\%)$ & $r (\%)$ & FP & Err ($\times 10^{-7}$) \\ \hline
			1 & 1 & 2.0 & 10 & 2 & 0.0559862 & 2 \hspace{0.5em} \\
			2 & 1 & 2.0 & 30 & 18 & 0.2183878 & 3 \hspace{0.5em} \\
			3 & 2 & 2.0 & 25 & 1.25 & 0.1722685 & $-$2 \hspace{0.5em} \\
			4 & 1 & 1.9 & 50 & 5 & 0.1931733 & $-$5 \hspace{0.5em} \\
			5 & 1 & 2.0 & 50 & 5 & 0.2464156 & $-$1 \hspace{0.5em} \\
			6 & 1 & 2.1 & 50 & 5 & 0.3062206 & 2 \hspace{0.5em} \\
			7 & 2 & 2.0 & 50 & 5 & 0.3500929 & $-$24 \hspace{0.5em} \\
			\hline
		\end{tabular}
	\end{center}
\end{table}

\section{Conclusion \label{sec:conc}}
Option pricing under a multivariate BSM model is a challenging task because of the curse of dimensionality, and has a long history of research. This study eases this curse significantly by replicating the option prices as a weighted sum of single-factor BSM prices under an optimally rotated state space.
Moreover, this method can be uniformly applied to spread, basket, and Asian options. Numerical examples show that this method is both fast and accurate.

\bibliographystyle{plainnat}
\bibliography{basket}

\end{document}